\documentstyle[aps,prl,epsfig,floats]{revtex}
\begin{document}
\twocolumn[\hsize\textwidth\columnwidth\hsize\csname
@twocolumnfalse\endcsname

\title{
Large Coulomb corrections to the $e^+e^-$ pair production
at relativistic heavy ion colliders
}
\author{
D. Yu. Ivanov\cite{DI}
}
\address{Institute of Mathematics, Novosibirsk, 630090, Russia}
\author{
A. Schiller\cite{AS}
}
\address{Institut f\"ur Theoretische Physik and NTZ,
          Universit\"at Leipzig, D-04109 Leipzig, Germany}
\author{
V. G. Serbo\cite{VS}
}
\address{Novosibirsk State University, Novosibirsk, 630090, Russia}
\date{November 30, 1998}

\maketitle

\vspace*{-5.7cm}
\noindent
\hfill \mbox{UL-NTZ 24/98, hep-ph/9809449}
\vspace*{5.3cm}

\begin{abstract}\noindent
We consider the Coulomb correction (CC) to the $e^+e^-$ pair
production related to multiphoton exchange of the produced
$e^{\pm}$ with nuclei. The contribution of CC to  the energy
distribution of $e^+$ and $e^-$ as well as to the total pair
production cross section are calculated with an accuracy of the
order of $1\, \%$. The found correction to the total Born cross
section is negative and equal to $-25\, \%$ at the RHIC for Au--Au
and $-14\, \%$ at the LHC for Pb--Pb collisions.
\end{abstract}

\vskip1.5pc]

\begin{figure}[!htb]
\begin{center}
\epsfig{file=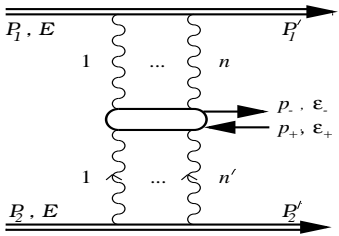,width=30mm,height=45mm,angle=270}
\vspace{5mm}
\caption{The amplitude $M_{nn'}$ of the process (2)
with $n$ $(n')$ virtual
photon emitted by the first (second) nucleus.}
\label{f1}
\end{center}
\end{figure}

\paragraph*{Introduction.} 
Two new large colliders with relativistic heavy nuclei, the RHIC
and the LHC, are scheduled to be in operation in the nearest
future. The charge numbers  $Z_1=Z_2=Z$ of the nuclei 
with masses $M_1=M_2=M$ and their
Lorentz factors $\gamma_1=\gamma_2=\gamma=E/M$ are the following
\begin{eqnarray}
Z=79\,, \ \gamma &=&\,\;108 \ {\mathrm {for \ RHIC \ (Au--Au \ collisions)}}\,
\nonumber \\
Z=82\,, \ \gamma &=&3000 \ {\mathrm {for \ LHC \ \ (Pb--Pb \ collisions)}}\,.
\label{1}
\end{eqnarray}
Here $E$ is the heavy ion energy in the c.m.s. 
One of the important processes at these colliders is
\begin{equation}
 Z_1Z _2\to Z_1Z_2 \, e^+e^- \,.
\label{2}
\end{equation}
Its cross section is huge. In the Born approximation (see
Fig.~\ref{f1} with $n=n'=1$) the total cross section according to
the Racah formula~\cite{R} is equal to
$\sigma_{\mathrm{Born}} = 36 $ kbarn for the RHIC and 227 kbarn
for the LHC. Therefore it will contribute as a serious background
to a number of experiments, besides, this process is the leading
beam loss mechanism (for details see review~\cite{BB}).

The cross sections of the process (\ref{2}) in the Born
approximation are known with accuracy $\sim 1/ \gamma^2$  (see,
for example, Refs.~\cite{R,KLBGMS} and more recent calculations
reviewed in Refs.~\cite{BB,BHT}). However, besides of the Born
amplitude $M_{\mathrm {Born}} =M_{11}$, also other amplitudes
$M_{nn'}$ (see Fig.~\ref{f1}) have to be taken into account for
heavy nuclei since in this case the parameter of the perturbation
series $Z\alpha$ is of the order of unity. Therefore, the  whole
series in $Z\alpha$ has to be summed to obtain the cross section
with sufficient accuracy. Following Ref.~\cite{BM}, we call the
Coulomb correction (CC) the difference $d\sigma_{\mathrm{Coul}}$
between the whole sum $d \sigma$ and the Born approximation
\begin{equation}
d\sigma = d \sigma_{\mathrm{Born}} + d \sigma_{\mathrm{Coul}}\,.
\label{4}
\end{equation}

Such kind of CC is well known in the photoproduction of $e^+e^-$
pairs on atoms (see Ref.~\cite{BM} and \S 98 of~\cite{BLP}). The
Coulomb correction to the total cross section of that process
decreases the Born contribution by about 10 \% for a Pb target.
For the pair production of reaction (\ref{2}) with $Z_1\alpha \ll
1$ and $ Z_2 \alpha \sim 1$  CC has been obtained in Refs.~\cite{NP,BB}.
Recently this correction has been calculated for the pair
production in the collisions of muons with heavy nuclei
\cite{IKSS}. The results of Refs.~\cite{NP,BB,IKSS} agree with
each other in the corresponding kinematic regions and noticeably
change the Born cross sections. Formulae for CC for two heavy
ions were suggested ad hoc in Sect. 7.3 of ~\cite{BB}.  However,
our calculations presented here do show that this suggestion is
incorrect.

In the present paper we calculate the Coulomb correction for
process (\ref{2}) omitting terms of the order of $1$ \% 
compared with the main term given by the Born cross section. We find
that these corrections are negative and quite important:
\begin{eqnarray}
\sigma_{\mathrm{Coul}}/ \sigma_{\mathrm{Born}} &=& -25\, \% \;\;
{\mathrm for \ \ RHIC}\,, \nonumber \\
\sigma_{\mathrm{Coul}}/ \sigma_{\mathrm{Born}} &=& -14\, \% \;\;
{\mathrm for \ \ LHC}\,.
\label{5}
\end{eqnarray}
This means that at the RHIC the background process with the
largest cross section will have a production rate 25 \% smaller
than expected.

Our main notations are given in Eq. (\ref{1}) and
Fig.~\ref{f1}, besides, $(P_1+P_2)^2 = 4E^2 = 4 \gamma^2 M^2$,
$q_i= (\omega_i,\, {\bf q}_i)= P_i-P_i'$, $\varepsilon=
\varepsilon_++\varepsilon_-$ and
\begin{equation}
\sigma_0=\frac{\alpha^4 Z_1^2 Z_2^2}{\pi m^2} \,, \;\;
L= \ln{P_1P_2 \over 2M_1 M_2}= \ln{\gamma^2}
\label{3}
\end{equation}
where $m$ is the electron mass. The quantities  ${\mathbf
q}_{i\perp}$ and ${\mathbf p}_{\pm\perp}$ denote the transverse
part of the corresponding three--momenta. Throughout the paper we
use the well known function\cite{BM}
\begin{equation}
f(Z) = Z^2\alpha^2 \sum_{n=1}^{\infty}
{1\over n(n^2+Z^2\alpha^2)}\,,
\end{equation}
its particular values for the colliders under discussion are
$f(79)=0.313$ and $f(82)=0.332$.

\paragraph*{Selection of the leading diagrams and the structure of
the amplitude.} 
Let ${\cal M}$ be the sum of the amplitudes $M_{nn'}$ of
Fig.~\ref{f1}.  It can be presented in the form
\begin{eqnarray}
\label{7a}
{\cal M}&=& \sum_{nn'\geq 1 } M_{nn'}= M_{\mathrm{Born}}
+M_1+{\tilde M}_1+ M_2\,,\\
M_1 &=& \sum_{n'\geq 2} M_{1n'}\,, \ \
\tilde M_1 = \sum_{n\geq 2} M_{n1}\,, \ \
M_2= \sum_{nn'\geq 2} M_{nn'} \,. \nonumber
\end{eqnarray}
The Born amplitude $M_{\mathrm{Born}}$ contains the one--photon
exchange both with the first and the second nucleus, whereas the
amplitude $M_1$ ($\tilde M_1$) contains the one--photon exchange
only with the upper (lower) nucleus. In the last amplitude $M_2$
we have no one--photon exchange. According to this
classification we write the total cross section as
\begin{equation}
\sigma = \sigma_{\mathrm{Born}} +\sigma_1 +\tilde\sigma_1 + \sigma_2
\label{7}
\end{equation}
where
\begin{eqnarray}
&&d\sigma_{\mathrm{Born}} \propto |M_{\mathrm{Born}}|^2\,,
\nonumber \\
&&d\sigma_1 \propto 2 {\mathrm Re}(M_{\mathrm{Born}} M_1^*) +|M_1|^2 \,,
\nonumber \\
&&d\tilde\sigma_1 \propto 2 {\mathrm Re}(M_{\mathrm{Born}} \tilde M_1^*) +|
\tilde M_1|^2 \,,
\nonumber \\
&&d\sigma_2 \propto 2 {\mathrm Re}\left( M_{\mathrm{Born}} M_2^* +
M_1\tilde M_1^* +M_1M_2^* \right.
\nonumber \\
 && \left. \hspace{1cm}+\tilde M_1M_2^* \right) + |M_2|^2 \,.
\nonumber
\end{eqnarray}

It is not difficult to show that the ratio $\sigma_i /
\sigma_{\mathrm{Born}}$ is a function of $(Z\alpha)^2$ only but
not of $Z \alpha$ itself. Additionally we estimate the leading
logarithms appearing in the cross sections $\sigma_i$.
The integration over the transfered momentum squared $q_1^2 $ and
$q_2^2$ results in two large Weizs\"acker--Williams (WW)
logarithms $\sim L^2$ for the $\sigma_{\mathrm{Born}}$, in one
large WW logarithm $\sim L$ for $\sigma_1$ and
$\tilde\sigma_1$. The cross section $\sigma_2$ contains no large
WW logarithm. Therefore, the relative contribution of the cross
sections $\sigma_i$ is $\sigma_1 / \sigma_{\mathrm{Born }}
=\tilde\sigma_1 / \sigma_{\mathrm{Born}} \sim (Z\alpha)^2 /L$ and
${\sigma_2 / \sigma_{\mathrm{Born}}} \sim (Z\alpha)^2 /L^2 \, <
0.4$\, \%  for the colliders (\ref{1}). As a result, with an
accuracy of the order of $1 \%$  we can neglect $\sigma_2$ in the
total cross section and use the equation
\begin{equation}
\sigma =  \sigma_{\mathrm{Born}} +\sigma_1 +\tilde\sigma_1\,.
\label{9}
\end{equation}
With that accuracy it is sufficient to calculate $\sigma_1$ and
$\tilde\sigma_1$ in the leading logarithmic approximation (LLA)
only since the next to leading log terms are of the order of
$(Z\alpha /L)^2$. This fact greatly simplifies the calculations.

The calculation in the LLA can be performed using the equivalent
photon or WW approximation. The main contribution to $\sigma_1$
and $\tilde\sigma_1$ is given by the region $(\omega_1/\gamma)^2
\ll -q_1^2 \ll m^2$ and $(\omega_2/\gamma)^2 \ll -q_2^2 \ll m^2$,
respectively.  In the first region the main contribution arises
from the amplitudes $M_{\mathrm{Born}}+M_1$
(in the second region $M_{\mathrm{Born}}+\tilde M_1$). The virtual photon
with four--momentum $q_1$ is almost real and the amplitude can be
expressed via the amplitude $M_\gamma$ for the real
photoproduction $\gamma Z_2\to Z_2 e^+e^-$ (see, for example, \S
99 of Ref.~\cite{BLP})
\begin{equation}
M_{\mathrm{Born}}+M_1\approx \sqrt{4 \pi \alpha} Z_1 \frac{
|{\mathbf{q}}_{1\perp}|} {(-q_1^2)} \, \frac{2 E}{\omega_1} \,
M_\gamma \,.
\label{amp}
\end{equation}
The amplitude $M_\gamma$ has been calculated in Ref.~\cite{BM}.
We use the convenient form of that amplitude derived in the works \cite{OM}
and \cite{IM}:
\begin{equation}
M_\gamma= ( f_1 \, M_\gamma^{\mathrm{Born}} + {\mathrm i} f_2 \,
\Delta M_\gamma) \, {\mathrm e}^{{\mathrm i}  \Phi}
\label{Mgamma}
\end{equation}
where $M_\gamma^{\mathrm{Born}}$ is the Born amplitude for the
$\gamma Z_2 \to Z_2 e^+ e^-$ process. This Born amplitude depends
on the transverse momenta ${\mathbf p}_{\pm\perp}$ only via the
two combinations $A=\xi_+-\xi_-$ and ${\mathbf B}=\xi_+ {\mathbf
p}_{+\perp} +  \xi_- {\mathbf p}_{-\perp}$ where $\xi_\pm= m^2/(
m^2+{\mathbf p}_{\pm\perp}^2)$. The quantity $\Delta M_\gamma$
is obtained from $M_\gamma^{\mathrm{Born}}$  replacing
$A\to \xi_++\xi_--1$ and ${\mathbf B}\to \xi_+ {\mathbf
p}_{+\perp} - \xi_- {\mathbf p}_{-\perp}$.

All the nontrivial dependence on the parameter $Z_2 \alpha \equiv
\nu$ are accumulated in the Bethe-Maximon phase
\begin{equation}
\Phi=\nu \, \ln\frac{(p_+P_2) \xi_+}{(p_-P_2) \xi_-}
\label{phase}
\end{equation}
and in the two functions (with $z=1 - (-q_2^2/m^2) \xi_+\xi_-$)
\begin{equation}
f_1=\frac{ F({\mathrm i} \nu,-{\mathrm i} \nu; 1 ; z)}
{ F({\mathrm i} \nu,-{\mathrm i} \nu; 1 ; 1 )} \,,\ \
f_2=\frac{1-z}{\nu} f_1'(z)\,.
\label{f1f2}
\end{equation}
The function $f_1(z)$ and its derivative $f_1'(z)$ are given
with the help of the Gauss hypergeometric function $F(a,b;c;z)$.

It can be clearly seen that in the region ${\mathbf
p}_{\pm\perp}^2 \sim m^2$ the amplitude $M_\gamma$ differs
considerably from the $M_\gamma^{\mathrm{Born}}$ amplitude and,
therefore, the whole amplitude ${\cal M}$ differs from its Born
limit $M_{\mathrm{Born}}$.  Let us stress that just this
transverse momentum region ${\mathbf p}_{\pm\perp}^2 \sim m^2$
gives the main contribution into the total Born cross section
$\sigma_{\mathrm{Born}}$ and into $\sigma_1$.

Outside this region the CC vanishes. Indeed, for ${\mathbf
p}_{\pm\perp}^2 \ll m^2$ or ${\mathbf p}_{\pm\perp}^2 \gg m^2$
the variable $ z \approx 1$, therefore, $f_1\approx 1$, $f_2
\approx 0$ and
\begin{equation}
M_{\mathrm{Born}}+M_1= M_{\mathrm{Born}} {\mathrm e}^{{\mathrm i} \Phi} \,.
\label{limit}
\end{equation}
Note that the region ${\mathbf p}_{\pm\perp}^2 \gg m^2$ gives a
negligible contribution to the total cross section $\sigma$,
however, this region might be of interest for some experiments.

The results of Ref.~\cite{BM} which are used here in the form of
Eqs.~(\ref{amp})-(\ref{limit}) are the basis  for our
consideration. These results were confirmed in a number of
papers (see, for example, Refs.~\cite{Qclas,IM}) using 
various approaches. 

Recently in Refs.~\cite{DIRACEQ}
the Coulomb effects were studied within the frame--work of a light--cone or
an eikonal approach. However, the approximations used in
Refs.~\cite{DIRACEQ} fail to reproduce the classical results of
Bethe and Maximon \cite{BM}. To show this explicitly, we
consider the simple case $Z_1 \alpha \ll 1, \; Z_2 \alpha \equiv
\nu \sim 1$ in which the principal result of Refs.~\cite{DIRACEQ}
for the amplitude takes the form 
${\cal M}=M_{\mathrm{Born}}+ M_1= M_{\mathrm{Born}} \exp
({\mathrm i} \Psi)$ with $\Psi = \nu \ln {\mathbf q}_{2\perp}^2$
in obvious contradiction to Eqs.  (\ref{amp})-(\ref{Mgamma}).
Since in the works \cite{DIRACEQ} different statements on the
applicability range of their results can be found, we take as an
example the common region $(\omega_i/\gamma_i)^2 \ll {\mathbf
q}_{i\perp}^2 \ll m^2$. But even in that region their expression
for the matrix element does not reproduce Eq.~(\ref{limit}) since
their phase $\Psi$ does not coincide with the Bethe--Maximon
phase $\Phi$, i.e. $\Psi\neq \Phi$.

\paragraph*{CC to the energy distribution and to the  total cross
section.} 
As it was explained in the previous section, the basic expression
for the cross section  $d\sigma_1$ in the LLA can be directly
obtained using the WW approximation. To show clearly the
terms omitted in the LLA, we start with a more exact expression
for $d \sigma_1$ derived for the case of $\mu Z $ collisions
considered in Ref.~\cite{IKSS}. The reason is that for the most
interesting region (when the energy of relativistic $e^{\pm}$
pairs is much smaller than the nucleus energy) the muon in the
$\mu Z$ scattering as well as the upper nucleus of the ion--ion
collision can be equally well treated  as spinless and pointlike
particles.

Using Eqs.~(14) and (17) from Ref.~\cite{IKSS} (given in the lab
frame of the muon projectile on a nucleus target)
and the invariant variables $x_{\pm}= (p_{\pm} P_2)/(q_1 P_2)$,
$y= (q_1 P_2) / (P_1P_2)$ we obtain $d \sigma_1$
for the pair production in $Z_1Z_2$ collisions
in the invariant form (and at $y\ll 1$)

\begin{eqnarray}
d\sigma_1& =&
- \frac{4}{3} \sigma_0 f(Z_2)
\left\{
\left[ (1+\xi) a-1\right] \ln \frac{1+\xi}{\xi} - \right.
\nonumber \\
&-& \left. a + \frac{4-a}{1+\xi} \, \right\}
{dy\over y} dx_+dx_- \delta(x_++x_- - 1)
\label{10}
\end{eqnarray}
with $a= 2 (1+x_+^2+x_-^2) \,, \ \
\xi= \left( M_1 y/m\right)^2 x_+x_- $.
The main contribution to $\sigma_1$ is given by the region
\begin{equation}
\frac{M_1^2 M_2^2 }{(P_1 P_2)^2} \ll \xi \ll 1\,.
\label{11}
\end{equation}
The corresponding expression for $d\tilde\sigma_1$ can be obtained by
making the replacements
\begin{equation}
d\tilde\sigma_1 = d\sigma_1(q_1 \to q_2, P_1 \leftrightarrow P_2,
Z_1\leftrightarrow Z_2)\,.
\label{12}
\end{equation}

Below we consider only the experimentally most interesting case
when in the collider system ($\gamma_1=E_1/M_1 \sim
\gamma_2=E_2/M_2$) both $e^+$ and $e^-$ are ultrarelativistic
($\varepsilon_\pm \gg m$). We assume that the $z$-axis is
directed along the initial three-momentum of the first nucleus
${\mathbf P}_1$.

To obtain the energy distribution of $e^+$ and $e^-$ in the LLA
we have to take into account two regions $p_{\pm z} \gg m$ and
$(-p_{\pm z}) \gg m$
where the
lepton pair is produced either in forward or backward direction.
In the first region we have $ x_\pm=
\varepsilon_{\pm}/\varepsilon$, $y=\varepsilon/E_1$, and from Eq.
(\ref{10})-(\ref{11}) we obtain  in the LLA
\begin{eqnarray}
&d\sigma_1^{(1)}&=
\nonumber \\
&-&4 \,\sigma_0  f(Z_2)
\left(1 - \frac{4\varepsilon_+  \varepsilon_-}{3 \varepsilon^2}
\right) \, \ln \frac{(m \gamma_1)^2}{ \varepsilon_+ \varepsilon_-}
 \, {d\varepsilon_+ d \varepsilon_-
\over \varepsilon^2} \,,
\label{13new} \\
&& m \ll \varepsilon_\pm \ll m \gamma_1 \,.
\nonumber
\end{eqnarray}
In the second region we have $x_\pm \approx \varepsilon_{\mp}/
\varepsilon$, $y\approx m^2 \varepsilon /( 4 E_1
\varepsilon_+\varepsilon_-)$) and
\begin{eqnarray}
&d\sigma_1^{(2)}&=
\nonumber \\
&-&4 \,\sigma_0  f(Z_2)
\left(1 - \frac{4\varepsilon_+  \varepsilon_-}{3 \varepsilon^2}
\right) \, \ln \frac{\gamma_1^2 \varepsilon_+ \varepsilon_-}{m^2}
 \, {d\varepsilon_+ d \varepsilon_-
\over \varepsilon^2} \,,
\label{14new} \\
&& m \ll \varepsilon_\pm \ll m \gamma_2 \,.
\nonumber
\end{eqnarray}
Summing up these two contributions, we find
\begin{equation}
d\sigma_1= - 8 \, \sigma_0  f(Z_2)
\left(1 - \frac{4\varepsilon_+  \varepsilon_-}{3 \varepsilon^2}
\right) \, \ln \gamma_1^2 \, {d\varepsilon_+ d \varepsilon_-
\over \varepsilon^2} \,.
\label{13}
\end{equation}

To obtain $\sigma_1$ we have to integrate the expressions
(\ref{13new}) and (\ref{14new}) over $\varepsilon_-$ (with
logarithmic accuracy)
\begin{eqnarray}
d\sigma_1^{(1)} &=& - \frac{28}{9} \sigma_0  f(Z_2)
 \, \ln \frac{(m \gamma_1)^2}{\varepsilon_+^2} \, \frac{d\varepsilon_+}
{\varepsilon_+} \,,
\label{15new} \\
&&m\ll \varepsilon_+\ll m \gamma_1 \,,
\nonumber
\end{eqnarray}
\begin{eqnarray}
d\sigma_1^{(2)} &=& - \frac{28}{9} \sigma_0  f(Z_2)
 \, \ln \frac{(\gamma_1 \varepsilon_+)^2}{m^2} \, \frac{d\varepsilon_+}
{\varepsilon_+} \,,
 \label{16new} \\
&&m\ll \varepsilon_+\ll m \gamma_2
\nonumber
\end{eqnarray}
from which it follows that
\begin{equation}
d\sigma_1= - \frac{28}{9} \sigma_0  f(Z_2)\,
\ln \gamma_1^2 \, \frac{d\varepsilon_+}
{\varepsilon_+} \,.
\label{17new}
\end{equation}
The further integration of Eqs. (\ref{15new}), (\ref{16new}) over
$\varepsilon_+$
results in
\begin{equation}
\sigma_1=- \frac{28}{9} \sigma_0  f(Z_2)\,
\left[ \ln \frac{P_1 P_2}{ 2 M_1 M_2} \right]^2 \,.
\label{18new}
\end{equation}
This expression is in agreement with the similar result
for the $\mu Z$ scattering (see Eq. (31) from \cite{IKSS}
for $Z_1 =1, Z_2=Z$).

The corresponding formulae for $\tilde\sigma_1$ can be obtained from
Eqs. (\ref{13}), (\ref{17new}) and  (\ref{18new}) by replacing
$\gamma_1\leftrightarrow \gamma_2$, $Z_1\leftrightarrow Z_2$.
The whole CC contribution $d \sigma_{\mathrm{Coul}}= d( \sigma_1+
\tilde\sigma_1)$ for the symmetric case $Z_1=Z_2=Z$ and
$\gamma_1=\gamma_2=\gamma$ takes the following form
\begin{equation}
d\sigma_{\mathrm Coul}= - 16 \,\sigma_0  f(Z)
\left(1 - \frac{4\varepsilon_+  \varepsilon_-}{3 \varepsilon^2}
\right) \, L \, \frac{d\varepsilon_+ d \varepsilon_-}
{\varepsilon^2}
\label{133}
\end{equation}
at $ m\ll \varepsilon_{\pm} \ll m \gamma \,$,
\begin{equation}
d\sigma_{\mathrm Coul}= - \frac{112}{9} \sigma_0  f(Z)
 \, L\, \frac {d\varepsilon_+ }
{\varepsilon_+}
\label{1333} \\
\end{equation}
at $ m\ll \varepsilon_+ \ll m \gamma \,$, and
\begin{equation}
\sigma_{\mathrm Coul}=- \frac{56}{9} \sigma_0  f(Z)\,
L^2 \,.
\label{188new}
\end{equation}

\begin{figure}[!htb]
\begin{center}
\epsfig{file=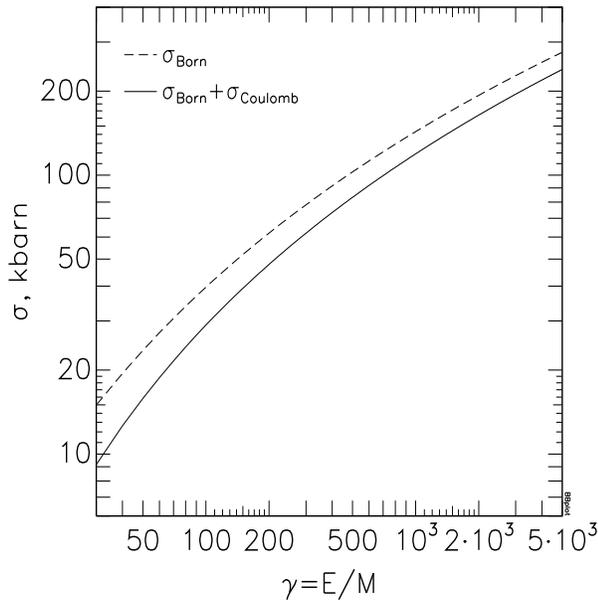,width=80mm,height=80mm}
\vspace{5mm}
\caption{The total cross section of the process
$ZZ \to$ $ZZe^+e^-$ with (solid line)  and without
(dashed line) Coulomb correction
as function of   the Lorentz factor
$\gamma$ of Pb nuclei ($Z=82$).}
\label{f3}
\end{center}
\end{figure}

The size of this correction for the two colliders was given before in
Eq. (\ref{5}). The total cross section with and without Coulomb correction as
function of the Lorentz factor $\gamma$ is illustrated in Fig.~\ref{f3} for
Pb nuclei.

\paragraph*{Conclusion.} 
We have calculated the Coulomb corrections to $e^+e^-$ pair
production in relativistic heavy ion collisions for the case of
colliding beams. Our main results are given in Eqs.
(\ref{133})-(\ref{188new}). We have restricted ourselves to the
Coulomb corrections for the energy distribution of electrons and
positrons and for the total cross section. 
In our analysis we neglected  contributions which are of the
relative order of $\sim (Z\alpha)^2/L^2$.
The CC to the angular
distribution of $e^+e^-$ can be easily obtained in a similar way, however 
only with an accuracy $Z\alpha/L^2$.

Since our basic formulae (\ref{10}), (\ref{12}) are given in the
invariant form, a similar calculation can be easily repeated for
fixed--target experiments.  This interesting question will be
considered in a future work.

{\it Acknowledgments.} ---
We are very grateful to G.~Baur, Yu.~Dokshitzer, U.~Eichmann, V.~Fadin,
I.~Ginzburg and V.~Telnov for useful discussions.  V.G.S.
acknowledges support from Volkswagen Stiftung (Az. No. I/72 302).
D.Yu.I. and V.G.S. are partially supported by the Russian
Foundation for Basic Research (code 96-02-19114).

\end{document}